\documentstyle[aps,preprint,prb]{revtex}

\begin{document}

\title{The Zeno effect and an inter-layer pairing mechanism for high-temperature
superconductivity in layered materials}

\author{N. Kumar}

\address{Raman Research Institute, Bangalore 560080, India\\
E-mail: nkumar@rri.ernet.in}

\maketitle

\begin{abstract}
 Quantum Zeno Effect (QZE) is the suppression of the inter-subspace
transition by a relatively fast intra-subspace decoherence.  Earlier, we had
proposed a QZE-based mechanism for the temperature-dependent normal-state c-axis
resistivity of the layered high-T$_c$ cuprate superconductors in which the
single-particle inter-layer tunneling is blocked by the strong intra-layer
decoherence (entanglement).  We now argue that while the single-particle
inter-layer tunneling is thus blocked, the tunneling of the bosonic BCS-like
pairs must remain unblocked inasmuch as a BCS pairing condensate is an
eigenstate of the pair annihilation operator.  This pair tunneling stabilizes
high-T$_c$ superconductivity energetically.
\end{abstract}

\section{Introduction} 
At the very outset, let me state the general point of
this work, which is that the normal-state c-axis resistivity is implicated in
the high-temperature superconductivity of the layered cuprates.  These
high-T$_c$ superconductors are by now well known to be qualitatively anisotropic
in their normal-state kinetic properties.$^{1-3}$ The qualitative anisotropy is
most evident in their normal-state electrical resistivities $-$ the out-of-plane
resistivity $\rho_c(T)$ along the c-axis and the in-plane resistivity
$\rho_{ab}(T)$ in the ab-plane $-$ both in terms of their absolute magnitudes
and the temperature dependences.  Thus, $\rho_c(T)$ far exceeds the
Mott-Ioffe-Regel (MIR) maximum metallic resistivity, with a transport mean-free
path that works out to be much smaller than the inter-planar spacing.  Still, it
has a positive temperature coefficient of resistance ($TCR > 0$) at high
temperatures, but develops a semiconductor-like resistive upturn as one
approaches the T$_c$.  The $\rho_{ab}(T)$, on the other hand , has a sub-MIR
metallic value with a positive TCR, and remains linear in $T$ down to about
T$_c$.  So, the in-plane conduction is band-like, or relatively coherent (i.e.,
there are propagating, although much scattered, Bloch waves).  In sharp contrast
to this the c-axis transport is of the hopping-type, or incoherent (the
successive inter planar tunnelings are not phase correlated and one may not
speak of the Bloch waves).  The individual inter-layer tunneling is, however,
phase coherent.  Most importantly, at high temperatures, the c-axis resistivity
tracks the ab-plane resistivity, i.e., $\Delta \rho_c(T) \propto
\Delta\rho_{ab}(T)$ down to a temperature that decreases towards $T_c$ as the
doping increases towards the optimal value corresponding to a maximum of $T_c$
within the given family.  Indeed, for the optimally doped, high-quality
untwinned single crystals, the T-linear behaviour of $\rho_c(T)$ as well as that
of $\rho_{ab}(T)$ has been reported to persist almost down to $T_c$.$^{4-5}$
This qualitative ly anisotropic electrical transport essentially involves the
Non-Fermi Liquid (NFL) nature of the strongly correlated electronic system at
hand, namely of the stacks of 2D CuO$_2$ sheets in the layered
cuprates.$^{3,6-8}$ The NFL behaviour is probed and revealed most clearly in the
angle-resolved-photoemission spectra (ARPES)$^9$ either by the striking absence
of any sharp quasi particle peaks in the normal state (which, however, do
reappear below the T$_c$) or by the presence of a large incoherent tail to the
quasi-particle peak as one moves away below the Fermi level.  The latter is
identifiable by the total disappearance of the quasi-particle peak.  The near
absence of the quasiparticle peak implies vanishing of the quasi-particle weight
\(Z_k \equiv (1 - \partial Re \Sigma_k(w)/\partial w)_{w=\epsilon_k}^{-1}\) as
$k\rightarrow k_F$, i.e., strong damping of the quasiparticle.  This loss of
coherence, and the consequent breakdown of the Landau FL picture, is also borne
out by the optical conductivity in that the transport scattering rate is found
to equal the quasi-particle energy.  This forms the basis of our recent$^{6,7}$
and the present work.  We first demonstrate that this intra-planar NFL dynamics
blocks the single-particle tunneling between the weakly coupled ab-planes.$^6$
Inasmuch as this suppression of the single-particle inter-planar tunneling may
be viewed as blocking of the inter-subspace transitions due to intra-subspace
coupling to the many-body environment (entanglement with the other-electronic
degrees of freedom), we refer to our blocking mechanism as due to the Quantum
Zeno Effect.$^{10-12}$ Next we argue, admittedly heuristically, that this
blocking mechanism may not be effective against the inter-planar tunneling of
Cooper $pairs$ for the planes prepared in the BCS-type trial pairing state,
which, therefore, gets stabilized.\\

\section{Blocking of the single-particle interplanar tunneling:  The Quantum
Zeno Effect} 
First, consider a bilayer of two strongly-correlated NFLs, A and B,
coupled weakly through the tunneling Hamiltonian (in obvious notation):
\begin{equation}
 H_\perp = -t_\perp \sum_{k\sigma} (b_{k\sigma}^\dagger
a_{k\sigma} + h.c.), \dots 
\end{equation} with $t_\perp > 0$.  The tunneling
conserves the in-plane momentum.\\

Now, the change $\Delta E_o$ in the ground state energy of the bilayer due to
this inter-planar tunneling can at once be written down by use of the
Hellmann-Feynman charging theorem (For no change of symmetry):  
\begin{equation}
\Delta E_o \equiv E_o(t_\perp) - E_o(0) = \frac{2t_\perp}{\pi} \sum_{k\sigma}
\int_0^\infty dw \int_0^1 d\eta Im G_{\perp\eta}^R (k, w) , 
\end{equation}
 where $G_{\perp \eta}^R(k, w)$ is the zero-temperature retarded Green 
function along the c-axis.  Here $\Delta E_o(<0)$ gives the lowering of the 
ground state energy.\\

Ignoring any vertex corrections to the inter-layer tunneling, we can express
$G_{\perp\eta}^R(k, w)$ in terms of the intra-planar Green function
$G_\perp^R(k, w)$, obtaining
\begin{equation}
G_\perp^R (k, w) = \frac{\eta t_\perp(G_\parallel (k,w))^2}
{1-\eta^2t_\perp^2(G_\parallel(k, w))^2}
\end{equation}
Thus, in this approximation the NFL nature enters through the in-plane Green
function $G_\parallel^R(k_1w)$ only.  We now use the semi-phenomenologically
validated Marginal Fermi Liquid (MFL)$^{13}$ expression for
$G_\parallel^R(k_1w)$ which is known to be consistent with a whole range of
normal-state experimental facts on HTSC.  This at once enables us to evaluate
$\Delta E_o$ analytically.  We find that the energy lowering, so calculated, due
to the single particle tunneling (delocalization) along the c-axis sharply
decreases as the MFL parameter characterizing the strength of the
electron-electron interaction is increased.$^7$ This clearly demonstrates that
the intra-planar NFL character leads to an effective blocking of the
single-particle tunneling along the c-axis, i.e., the single-particle tunneling
matrix element $t_\perp$ is renormalized to a $t_{\perp eff}^{(1)} \ll t_\perp$.
This ties up neatly with the Caldeira-Leggett idea of dynamical c-axis
localization due to intra-planar environmental couplin g.$^{14}$ Note that the
condition for the validity of our earlier proposal, namely that $\Delta
\rho_c(T) \propto \Delta \rho_{ab}(T)$, now gets relaxed from $k_BT > t_\perp$
to $k_BT > t_{\perp eff}^{(1)}$.  This, therefore, extends its domain of
validity to much lower temperatures, obviating thus the concern expressed by
some workers.$^{15}$\\

\section{Unblocked c-axis pair-tunneling and superconductivity}

For this, we again consider the weakly coupled bilayer comprising the planes A
and B.  Let the two many-body planar sub-systems be prepared in a trial state,
namely, the BCS-like state $|\psi> = |A>|B>$ with
\begin{equation}
|A> = \Pi_k (u_k + v_k a_{k\uparrow}^\dagger a_{-k\downarrow}^\dagger) \equiv
e^{g\alpha^\dagger} |0>,
\end{equation}
and similarly for $|B>$.
Here 
\begin{eqnarray*}g\alpha^\dagger = \sum_k \left( \frac{u_k}{i_k}\right) a_{k\uparrow}^\dagger 
a_{-k \downarrow}^\dagger ,
\end{eqnarray*}
and similarly for $g\beta^\dagger$.  Thus $\alpha^\dagger(\alpha)$ and
$\beta^\dagger(\alpha)$ are the pair creation (annihilation) operators for the
two layers A and B respectively.  Now, recall that for a dilute electronic
system, and also, presumably, for a small pair-size (as is the case for the
optimal hole doping levels in question), the pair operators are essentially
bosonic and the coherent states $|A>, |B>$ are eigenstates of the pair
annihilation operators $\alpha, \beta$ respectively, with the eigenvalue $g$.
Thus, a pair tunneling Hamiltonian
\begin{equation}
H_{AB}^{(2)} = -t_\perp^{(2)} (\alpha^\dagger \beta + \beta^\dagger \alpha)
\end{equation}
would subtend a pair-tunneling amplitude along the c-axis
\begin{equation}
<\psi | H_{AB}^{(2)} |\psi> = -2t_\perp^{(2)} |g|^2
\end{equation}
This should imply that the pairs can tunnel adiabatically, a 
point missed in the other inter-layer pairing theories.$^{16,17}$ \\

Crucial to our argument is the point that while the pair-tunneling is admittedly
a two-step process ($t_\perp^{(2)}$) involving a virtual intermediate
single-particle tunneling ($t_\perp$), we still have $T_\perp^{(2)} \sim
t_\perp$.  This is so because the intermediate state is nearly degenerate with
the initial and the final states.  Thus, we have the tunneling matrix elements:
$t_\perp^{(2)} \sim t_\perp \gg t_{\perp eff}^{(1)}$ This motivates us to
considering a reduced Hamiltonian $h_{red}$:
\begin{equation}
h_{red} = h_a + h_b + h_{ab}^{(2)} + h_{ab}^{(1)}
\end{equation}
with
\begin{eqnarray}
h_a &=& \sum_{k,\sigma} \epsilon_k a_{k\sigma}^\dagger a_{k \sigma} + \frac{u}{2N} \sum a_{k'\uparrow}^\dagger a_{-k'\downarrow}^\dagger a_{-k\downarrow} a_{k\uparrow}\nonumber\\
h_b &=& \sum_{k,\sigma} \epsilon_k b_{k\sigma}^\dagger b_{k \sigma} + \frac{u}{2N} \sum_{k,k'} b_{k'\uparrow}^\dagger b_{-k'\downarrow}^\dagger b_{-k\downarrow} b_{k\uparrow}\nonumber\\
h_{ab}^{(2)} &=& - \sum_{k,\sigma} t_\perp^{(2)} (k) (b_{k\sigma}^\dagger  b_{-k-\sigma}^\dagger  a_{k\sigma}a_{-k-\sigma} + h.c.)\nonumber\\
h_{ab}^{(1)} &=& -\sum{k,\sigma}t_{\perp eff}^{(1)} (k) (b_{k\sigma}^\dagger  a_{k\sigma}  
+ h.c.)
\end{eqnarray}
with $u>0$ (repulsion).
Here the usual Hubbard term with a strong on-site repulsion (ultimately
responsible for the NFL nature) is replaced by a reduced {\em pairing} term
(i.e., one which is repulsive, but still maintains the pairing condition)
together with the pair tunneling term, and a residual, relatively suppressed
single-particle tunneling term.  Of course, for the cuprates all the
inter-planar tunneling matrix elements have the angular ({\bf k}) dependence
$\propto (\cos ak_x - \cos ak_y)^2$, and hence dominant in the (0, $\pm\pi$) and
($\pm\pi$, 0) directions.  This together with the short-ranged repulsion should
favour a d-wave singlet pairing, as indeed is known to be the case.

The reduced Hamiltonian clearly supports a superconducting phase, at least at
the mean field level.  In order to see this in its essentiality, let us ignore
the angular dependence of the tunneling matrix elements, and just look for an
s-wave paired BCS-like state.  Introducing the anomalous average $\Delta$
(assumed real):
\begin{equation}
\Delta = \sum_k <a_{-k\downarrow} a_{k\uparrow}> = \sum_k <b_{-k\downarrow} b_{k\uparrow}>,
\end{equation}
the pair tunneling term reduces to
\begin{equation}
h_{ab}^{(2)} = - \left (\frac{t_\perp^{(2)}\Delta}{N}\right ) \sum_{\sigma,k} (b_{k\sigma}^\dagger a_{-k-\sigma} + h.c.),
\end{equation}
and similarly for the reduced Coulomb term containing u.\\

The resulting bilinear Hamiltonian can now at once be diagonalized through a
Bogoliubov transformation leading to a self-consistent gap equation for
$\Delta$.  Thus, for the simple case of $t_{\perp eff}^{(1)}$ = 0, we get for
the superconducting transition temperature $T_c$ (with $\gamma$ the Euler
constant, $v$ the density of states per spin, and $W \sim$ hole Fermi-energy):
\begin{equation}
k_BT_c = (4\gamma/\pi) W \exp(-2/\nu(t_\perp^{(2)}-u)),
\end{equation}

\section{Concluding Remarks}

I shall now conclude with a number of general remarks.  The inter-layer pairing
mechanism described above invokes the 2D-NFL (e.g., MFL) character for blocking
the single-particle c-axis tunneling through entanglement with the intra-layer
degrees of freedom $-$ the Quantum Zeno Effect.  A quantum liquid with spin
charge separation (e.g., a 2D Luttinger Liquid) seems to represent from our
point of view an extreme case of such a blocking (confinement).$^{3,15,18}$ \\

The present mechanism, however, must be fully confronted with the pseudo-gap
observed in the underdoped cuprates below a cross-over temperature $T^*(> T_c)$,
with $T^*$ decreasing with increasing hole doping.$^{19}$ The HTSC indeed seems
to be a God of Gaps!  More specifically, the strong temperature dependence of
the c-axis spectral weight$^{20}$ remains a puzzle.  (The conductivity sum rule
should be, of course, invariant under pairing, i.e., charge doubling mass
doubling and electron-number halving, and temperature-insensitive for a proper
choice of the frequency cut-off.) \\

Finally, we suggest that such qualitatively anisotropic materials exhibiting the
Zeno Effective blocking of the c-axis single-particle tunneling be aptly called
Zenophilic, and that the conditions for Zenophilicity, involving, e.g., the
tunneling time, the life time and the Zeno time be examined carefully.$^{21}$ \\

\end{document}